\documentclass[12pt]{spieman}  % 12pt font required by SPIE;
\usepackage{amsmath,amsfonts,amssymb}
\usepackage{graphicx}
\usepackage{setspace}
\usepackage{tocloft}
\usepackage{lineno}
\usepackage{xspace}

\newcommand{\dlux}{\textsc{$\partial$Lux}\xspace}
\newcommand{\jax}{\textsc{Jax}\xspace}

\newcommand{\equinox}{\textsc{Equinox}\xspace}
\usepackage{amsmath} % for \DeclareMathOperator
\DeclareMathOperator{\jinc}{jinc}

\title{Spectral Information Content of Diffraction-Limited Images}

\author[a,*]{\href{https://orcid.org/0009-0000-9575-6190}{Milo~Langker}}
\author[a]{\href{https://orcid.org/0009-0008-8188-8668}{Hayden~Greer}}
\author[b,c]{\href{https://orcid.org/0000-0003-2595-9114}{Benjamin~Pope}}
\author[a]{\href{https://orcid.org/0000-0001-7026-6291}{Peter~Tuthill}}
\affil[a]{Sydney Institute for Astronomy, School of Physics, University of Sydney, Camperdown, NSW 2006, Australia}
\affil[b]{School of Mathematical and Physical Sciences, Macquarie University, Sydney, NSW 2109, Australia}
\affil[c]{Astrophysics and Space Technologies Research Centre, Macquarie University, Sydney, NSW 2109, Australia}

\cftpagenumbersoff{figure}
\cftpagenumbersoff{table} 

\begin{document} 
\maketitle

\begin{abstract}
High-resolution optical imaging is fundamentally limited by diffraction, which imprints wavelength-dependent structure onto the point spread function (PSF) of the optical system. This intrinsic chromaticity implies that in addition to spatial structure, images encode spectral information. We develop the theory of this spectral information content and provide strategies for its optimal extraction with calculations of Fisher information. We show that polychromatic PSF formation can be expressed as a convolution in logarithmic coordinates and define the corresponding spectral transfer function. We derive this function for the Airy disk and find algebraic decay in spectral log-frequency, implying the existence of arbitrarily fine spectral structure in the ideal, noiseless PSF. We use a differentiable forward model to numerically construct the spectral information eigenbasis and quantify the spectroscopic capability of  diffraction-limited imaging across a range of noise and detector systematics. We demonstrate injection and recovery of the constrained spectral modes in a realistic simulated example and compare these results with previously reported methods. Our information-theoretic framework provides a generalisable method for spectral characterisation of PSFs spanning a variety of imaging contexts.
\end{abstract}

% Include a list of up to six keywords after the abstract

% Include email contact information for corresponding author
{\noindent \footnotesize\textbf{*}Author for correspondence: Milo Langker, \linkable{milo.langker@sydney.edu.au}}

% for submission
% \begin{spacing}{2}   % use double spacing for rest of manuscript

\section{Introduction}\label{sec1}
% no subheadings
Imaging systems are fundamentally constrained by the range of spatial frequencies they transmit, producing a diffraction-limited point spread function (PSF) whose spatial scale depends on wavelength \cite{goodman2005}. Although this effect is general to image formation, astronomy provides a natural setting for its study and motivates the focus of this work. While conventional ground-based astronomical instruments contend with the resolution-degrading effects of atmospheric turbulence, performance of space telescopes and terrestrial systems equipped with sufficiently advanced adaptive optics can approach the classical diffraction limit \cite{adaptive_optics2012}. Here we use the term `diffraction limit' to refer specifically to the finite spatial-frequency support imposed by the telescope aperture, rather than any image-plane resolution criterion -- such as Rayleigh's -- which may often be overcome in practice \cite{maznev2017}.

In many cases, observed targets exhibit structure near or beyond this resolution scale, necessitating precise calibration of the PSF; parametric modelling, regularisation, or proscriptive priors are often required for such inference \cite{starck2002}. Much attention has therefore been given to wavefront distortions (including field-dependent aberrations), optical stability, and detector effects (e.g., Refs.~\citenum{liaudat2023, amigo2026}), while the trade-off between photon collection and bandwidth smearing is a familiar concern in interferometry \cite{bandwidth1999}. Despite these considerations, the spectral information content retained in arbitrary diffraction-limited imaging has, to our knowledge, yet to be examined in a rigorous information-theoretic manner. While practical extraction of some form of spectral information has been demonstrated, this work instead seeks to quantify the fundamental recovery limits.

Before addressing information content, it is necessary to understand image formation from scalar electric fields in paraxial optical systems. Following the standard treatment of Goodman \cite{goodman2005}, when monochromatic light of wavelength $\lambda$ is propagated to the focal plane, its electric field $E$ is given by the 2-dimensional Fourier transform of its aperture-plane field $\mathcal{A}$, known as Fraunhofer diffraction
\begin{align}
    E(\mathbf{r}) &= \frac{1}{\lambda} \int \mathcal{A}(\mathbf{r}_{\rm p}) \exp \left[ -\frac{2\pi i}{\lambda} \mathbf{r} \cdot \mathbf{r_{\rm p}} \right] d^2 \mathbf{r_{\rm p}}, \label{eq:focal_field_vec}
\end{align}
where $\mathbf{r}=(x,y)$ indicates the focal-plane angular coordinate and $\mathbf{r_{\rm p}} = (x_{\rm p}, y_{\rm p})$ the aperture-plane physical coordinate, in the same units as $\lambda$. Note that in general there may be quadratic phase prefactors in front of the integral, however, since we are chiefly concerned with the PSF -- defined by the squared amplitude of this field -- we will neglect such factors. 

Defining $\boldsymbol \xi \equiv \mathbf r / \lambda$, and $g(\boldsymbol \xi)$ as
\begin{align}
    g(\boldsymbol \xi) \equiv \left|\int \mathcal{A}(\mathbf{r_{\rm p}}) \exp \left[ -2\pi i  \boldsymbol \xi \cdot \mathbf{r_{\rm p}} \right] d^2 \mathbf{r_{\rm p}} \right|^2,
\end{align}
we can express the monochromatic PSF as
\begin{align}
    p(\mathbf{r}|\lambda) = \frac{1}{\lambda^2} g\left( \frac{\mathbf{r}}{\lambda}\right).\label{eq:monopsf_lhood}
\end{align}

Since electric fields of different wavelengths propagate incoherently for thermal sources, we can express the polychromatic PSF as an integral of this monochromatic PSF over the spectral photon-flux distribution $S$. We denote this similarly to the monochromatic case, expressing the polychromatic PSF as
\begin{align}
    p(\mathbf{r} | S) = \int_0^\infty S(\lambda) p(\mathbf{r}|\lambda) d\lambda.\label{eq:polypsf_integral}
\end{align}

Although these equations are limited to paraxial Fraunhofer propagation, our numerical methods (Sec.~\ref{sec:eigenbasis}) are generalisable to any aperture-plane wavelength-dependent phase or amplitude aberrations, so long as they are parametrised by a differentiable model. Arbitrary paraxial propagation follows from the Collins integral, with the differences including scaling and chromatic radial quadratic phase factors parameterised by the system's relevant ABCD matrix \cite{collins1970}. 

In this work, we show that there exists a `spectral transfer function', analogous to the spatial-frequency optical transfer function (OTF), for the transmittance of spectral frequency information in polychromatic PSFs. The support of such a function is of great interest; one would intuitively expect relatively minimal spectral information to be present for optical systems lacking a dedicated dispersive element. The grating equation for spectral resolution ($R = |m|N_{\rm g}$ where $N_{\rm g}$ is the number of grating lines illuminated in the aperture, with $m$ the order of diffraction) is difficult to interpret for apertures without explicit periodic structure \cite{loewen1997diffraction}. However, one might analogously consider a filled aperture as a single-line grating ($N_{\rm g}=1$) with the successive diffraction sidelobes (e.g. the Airy rings in the circular aperture case) corresponding to increasing orders $m$. We thus expect a spectral resolution scaling with the largest ring present in the field of view, given sufficiently large signal-to-noise. The properties of the spectral transfer function are treated in Sec.~\ref{sec:spectralconv} explicitly.

Prior works have examined several consequences of the chromatic PSF integral in Eq.~\eqref{eq:polypsf_integral}, demonstrating the existence of spectral information (e.g., Refs.~\citenum{subbarao1990, han2017}). Recovery of this information has been achieved to varying degrees in previous studies. In particular, Wang and Menon~\cite{wang2014a, wang2014b} developed an optical element engineered to best encode the spectrum for computational reconstruction, and demonstrated this with singular value decomposition of the system-response matrix and Tikhonov regularisation of the recovered spectrum. Similarly, Yang \textsl{et al.}~\cite{yang2015} formulated a simplified 10-hole design and implemented regularised spectral recovery with the broadband PSF. In microscopy, engineered PSFs have been produced to simultaneously recover position and colour of single point emitters \cite{shechtman2016}. Building on these ideas, Chen, Gu, and Liu~\cite{chen2024} demonstrated regularised spectral extraction on a simpler optical system consisting of a single arbitrarily shaped pinhole. With only knowledge of the profile of the monochromatic PSF, their method explicitly uses the wavelength scaling of the polychromatic PSF to reconstruct its spectrum, again with Tikhonov regularisation. 

In the astronomical context, chromatic effects on the PSF have often been treated as a systematic to correct for rather than information to constrain and recover (e.g., Ref.~\citenum{carlsten2018}). Lee \textsl{et al.}~\cite{lee2019} demonstrated that differential chromatic refraction by the atmosphere -- rather than the wavelength-scaling of the PSF -- within a single passband could be used to infer the spectral energy distribution of the underlying source. More recently, Berlfein \textsl{et al.}~\cite{berlfein2025} quantified these spectral effects on \textsl{Roman Space Telescope}-like weak lensing surveys.

Despite the prevailing focus on mitigation and calibration, some astronomical studies have attempted to exploit the inherent chromaticity in broadband PSFs to recover (typically some function of) the underlying source spectrum. Kuntzer \textsl{et al.}~\cite{kuntzer2016} developed a method training a neural network on PCA-reduced data to classify spectral classes (i.e. OBAFGKM classification, corresponding to the effective temperature scale of stellar spectroscopy) on simulated \textsl{Hubble} and \textsl{Euclid Space Telescope} images, presenting a typical error of half a spectral class. Similarly, Centofanti \textsl{et al.}~\cite{centofanti2025} -- motivated by the precise chromatic PSF calibration required for accurate modelling of gravitational lensing -- demonstrated that they could break the degeneracy between field-dependent PSF aberrations and stellar spectral class. In only selecting a discrete categorisation, these astronomical methods do not attempt to recover the underlying spectrum from broadband photometry. More recently however, Cao \textsl{et al.}~\cite{cao2026} investigated joint spectrum-wavefront recovery on simulated, laboratory, and \textsl{Webb} data with a differentiable forward model. While information-theoretic tools were used in comparing optical configurations, their method, like those in the optics literature, does not explicitly identify the recoverable spectral degrees of freedom, characterise their uncertainties, or demonstrate their information-optimal retrieval. These gaps motivate the present work.

To quantify the PSF-encoded spectral information content, we adopt a Fisher matrix formalism. From a Bayesian perspective broadly following the treatment of Wakefield\cite{wakefield2013} and Gelman\cite{gelman2013}, the observed Fisher information matrix $\mathcal{J}$ can be defined as the negative Hessian of the log likelihood $p(\mathbf{d}|\boldsymbol{\theta})$ with respect to the parameters $\boldsymbol{\theta}$
\begin{align}
    \mathcal{J} (\boldsymbol{\theta})  = - \nabla_{\boldsymbol{\theta}}^2 \log p (\mathbf{d} | \boldsymbol{\theta}).
\end{align}
Under the Laplace approximation and assuming negligible log-prior curvature over the posterior support, the inverse of the observed Fisher information gives the covariance matrix of the relevant Gaussian posterior. Notably, the observed Fisher information depends on a particular instantiation of the data $\mathbf{d}$. Taking the expectation under a set of true parameter values $\boldsymbol{\theta}$ defines the expected Fisher information $\mathcal{I}$ \cite{fisher1925}, given by
\begin{align}
    \mathcal{I} (\boldsymbol{\theta}) = \mathbb{E}_{\mathbf{d} \sim p(\cdot | \boldsymbol{\theta})} \left[ - \nabla_{\boldsymbol{\theta}}^2 \log p (\mathbf{d} | \boldsymbol{\theta}) \right]. \label{eq:fisher}
\end{align}
The expected Fisher information is thus a property of the underlying statistical model, independent of any single data realisation. For brevity, we will omit the term `Fisher' in subsequent references to these objects.

The large-sample validity of the Laplace approximation is justified by the Bernstein--von Mises theorem \cite{kasprzak2025}, which states that (under appropriate regularity conditions) with increasing data draws the posterior converges to a normal distribution with inverse covariance given by the expected information. In frequentist statistics, the expected information is known to dictate the precision of the normal distribution from which maximum likelihood estimates are drawn around these true values. Its inverse (known as the Cramér--Rao bound \cite{rao1945, cramer1946}) gives a lower bound on the covariance of unbiased estimators of the data-generating parameters around those true parameters, with maximum likelihood being one such (only asymptotically unbiased) estimator. 

In astronomy, Fisher-information approaches have been used extensively to quantify uncertainties and guide experimental design, where maximising expected information corresponds to minimising the expected posterior spread (e.g., Refs.~\citenum{mendez2014, lee99, haffert2023, taras2026, adam2026, dlux2}). Here, we decompose these expected and observed information matrices to determine the constrainable parameter directions, a familiar technique across astronomy (e.g., Refs.~\citenum{tegmark1997, hunterer2003truncate, amigo2026}). To evaluate these derivative matrices in large parameter spaces, accurate and efficient numerical methods are required; we therefore use \dlux \cite{dlux1}, a differentiable optical modelling library built on \jax \cite{jax} and \equinox \cite{kidger2021equinox}. These libraries leverage automatic differentiation (`autodiff') \cite{autodiff} to efficiently calculate gradients of arbitrarily complex functions. 
\begin{figure}
    \centering
    \makebox[\linewidth][c]{
        \includegraphics[width=5.25in]{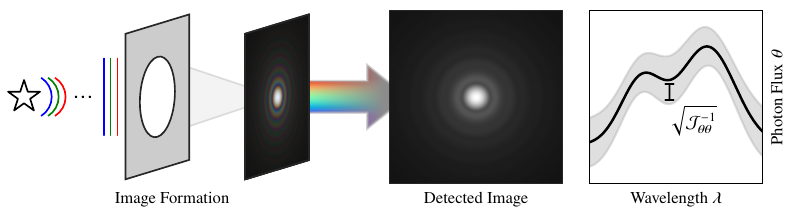}
    }
    \caption[Spectral information and recovery schema]{Illustration of the conceptual schema of this work, where we consider the spectral flux distribution of a polychromatic PSF recorded with a single broadband sensor. The inverse observed information gives the posterior variance under our assumptions of a negligible prior curvature.}
    \label{fig:intro_figure}
\end{figure}

In this work, we first construct a different factorisation of the polychromatic PSF integral in Eq.~\eqref{eq:polypsf_integral}, expressing it as an angle-dependent radial convolution in log-wavelength. We further derive the analytic form of the convolution kernel and hence the `spectral transfer function' for the Airy disk (Secs.~\ref{sec:spectralconv}/\ref{sec:airy_conv}). We then construct information matrices for various representative aperture geometries and project the number of recoverable modes, and thus an effective spectral resolution $R$, for a range of regimes (Secs.~\ref{sec:airy_example},~\ref{sec:fluxfov},~\ref{sec:read_sampling}, and~\ref{sec:jwst_hubble}). We finally demonstrate recovery of these modes and their relative uncertainties in the three relevant bases (Sec.~\ref{sec:injec_recov}).

\section{Results}\label{sec2}
We construct the spectral transfer function formalism of polychromatic PSF formation. We then consider the information content of the spectra of polychromatic PSFs with the information matrix formalism.

\subsection{Polychromaticity as Spectral Convolution}\label{sec:spectralconv}

\vspace{2ex}\noindent{\footnotesize\textbf{Proposition 1.} (Spectral convolution)
Suppose that the polychromatic PSF can be written as
\begin{align}
    p(r,\phi|S) = \int_0^\infty S(\lambda)\frac{1}{\lambda^2}g\left(\frac{r}{\lambda},\phi\right)d\lambda,\label{eq:poly_g_int}
\end{align}
where $r$ and $\phi$ are the radial and azimuthal focal-plane coordinates, $S(\lambda)$ is the normalised spectral photon-flux distribution, and the profile $g$ has no explicit wavelength dependence. Assume that $S$ is supported within a finite bandpass $0<\lambda_{\min}\leq\lambda\leq\lambda_{\max}<\infty$ and that $g_\phi(\xi)\equiv g(\xi,\phi)$ is integrable over $\xi\in(0,\infty)$ at the azimuths considered.

With logarithmic coordinates $\eta\equiv\log r$ and $\ell\equiv\log\lambda$, the PSF can also be written as a one-dimensional convolution at fixed azimuth $\phi$,
\begin{align}
    \Pi_\phi(\eta) = (\Sigma\ast K_\phi)(\eta),\label{eq:spectral_conv}
\end{align}
where
\begin{align}
    \Sigma(\ell)=S(e^\ell),\quad K_\phi(\eta)=e^\eta g(e^\eta,\phi),\quad \Pi_\phi(\eta)=e^\eta p(e^\eta,\phi|S).\label{eq:log_functions}
\end{align}
Equivalently, in the Fourier domain,
\begin{align}
    \widehat{\Pi}_\phi(\omega)=\widehat{\Sigma}(\omega)\cdot\widehat{K}_\phi(\omega),\label{eq:spectral_conv_fourier}
\end{align}
with spectral transfer function
\begin{align}
    \widehat{K}_\phi(\omega)=\mathcal{M}\{g_\phi\}(1-i\omega).\label{eq:mellin_psf_kernel}
\end{align}
}

Here the hat denotes the one-dimensional Fourier transform at angular frequency $\omega$ conjugate to the logarithmic coordinate, and $\mathcal{M}$ denotes the Mellin transform. Logarithms of dimensional quantities are understood to be logarithms of their numerical values in the adopted units. The transform conventions and proof are given in Methods, Sec.~\ref{sec:methods_spectralconv}.

By Eqs.~\eqref{eq:spectral_conv} and~\eqref{eq:spectral_conv_fourier} we have that there exists a spectral transfer function $\widehat{K}_\phi$, conditional on angular coordinate $\phi$, for polychromatic PSFs. Notably, this spectral transfer function operates on the log-frequency representation of the spectrum, reflecting that a change in wavelength imparts a multiplicative scaling on the corresponding monochromatic PSF. This logarithmic rescaling of the Fourier transform is a familiar technique in optical signal processing (e.g., Ref.~\citenum{casasent1976}); however, to our knowledge our use of it to transform the incoherent polychromatic integral into a convolution is novel. The magnitude of this spectral transfer function dictates how much the log-frequencies of the spectrum are attenuated by the PSF morphology.

There is a useful interpretation of the spectral log-frequency coordinate $\omega$. Noting that a Fourier mode of $\ell$ varies between adjacent extrema over half a period $\Delta\ell=\pi/|\omega|$, we can associate our angular frequency with an effective spectral resolution,
\begin{align}
    R_{\mathrm{eff.}}\equiv\frac{1}{2\sinh\left(\frac{\pi}{2|\omega|}\right)}\approx\frac{|\omega|}{\pi},\qquad |\omega|\gg1.\label{eq:log_resolution}
\end{align}
The conversion is derived in Sec.~\ref{sec:methods_spectralconv}.

While a generalised treatment of $K_\phi$ for PSFs from arbitrary apertures is beyond the scope of this work, we next examine the spectral transfer function of the Airy disk.

\subsubsection{Airy Disk Spectral Convolution Kernel}\label{sec:airy_conv}

For an Airy PSF generated by a clear, circular aperture of diameter $D$, the wavelength-independent profile is
\begin{align}
    g(\boldsymbol{\xi})=\pi D^2\jinc^2(\pi D\xi),\label{eq:airy_g}
\end{align}
where $\xi=\lVert\boldsymbol{\xi}\rVert$, $\boldsymbol{\xi}=\mathbf{r}/\lambda$, and $\jinc(x)\equiv J_1(x)/x$, with $J_1$ the Bessel function of the first kind of order 1. 

\vspace{2ex}\noindent{\footnotesize\textbf{Corollary 1.} (Airy disk spectral transfer function)
The spectral transfer function of the polychromatic Airy disk at arbitrary azimuth is
\begin{align}
    \widehat{K}(\omega)=\frac{\pi^{i\omega}D^{1+i\omega}}{2}\frac{\Gamma\left(1+\frac{i\omega}{2}\right)\Gamma\left(\frac{1}{2}-\frac{i\omega}{2}\right)}{\Gamma\left(\frac{1}{2}\right)\Gamma\left(\frac{3}{2}+\frac{i\omega}{2}\right)\Gamma\left(\frac{5}{2}+\frac{i\omega}{2}\right)},\label{eq:airy_spectral_kernel}
\end{align}
where $\Gamma(z)$ denotes the Gamma function. Its magnitude has the asymptotic form
\begin{align}
    |\widehat{K}(\omega)|\sim\frac{2\sqrt{2}D}{\sqrt{\pi}}|\omega|^{-5/2},\qquad |\omega|\rightarrow\infty.\label{eq:airy_kernel_asymptotic}
\end{align}
}

\begin{figure}
    \centering
    \makebox[\linewidth][c]{%
        \includegraphics[width=5.25in]{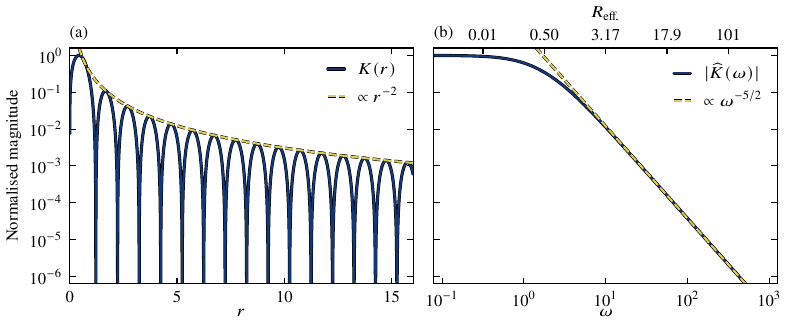}%
    }
    \caption[Airy spectral convolution kernel and transfer function]{The convolutional kernel for the polychromatic Airy disk in $r$ and its Fourier magnitude in $\omega$. (a) The normalised real-space kernel $K$ and its asymptotic envelope as a function of radius $r$. (b) The normalised magnitude of the Fourier kernel $\widehat{K}$ and its asymptotic as a function of spectral log-frequency $\omega$, with corresponding effective spectral resolution $R_{\mathrm{eff.}}$ also annotated.}
    \label{fig:airy_kernels}
\end{figure}

The proof is given in Methods, Sec.~\ref{sec:methods_airykernel}. The real-space convolutional kernel, its Fourier magnitude, and its asymptotics are displayed in Fig.~\ref{fig:airy_kernels}.

Rather than having finite support (as in the case of the optical transfer function for finite apertures), the spectral transfer function of the Airy disk is not band-limited, and instead asymptotes to a power law in spectral log-frequency. This algebraic attenuation implies that -- in principle -- arbitrarily fine spectral structure leaves some imprint on the polychromatic PSF, however in practice finite fields of view and noise constraints render recovery at higher spectral log-frequencies ill-conditioned. The extent to which these high log-frequency spectral modes can be recovered in the presence of these constraints is investigated in the following section with an information-theoretic approach.

\subsection{The Spectral Information Eigenbasis}\label{sec:eigenbasis}

To quantify which spectral modes remain constrainable under finite fields of view, noise, and detector sampling, we take a numerical approach to computing the spectral information content in polychromatic PSFs. We calculate the expected information matrix $\mathcal{I}$ with respect to the fluxes associated with each propagated wavelength, denoted by $\boldsymbol{\theta}$, for a specified spectrum, optical model, detector model, and likelihood. As $\mathcal{I}$ is real and symmetric, it has the following eigendecomposition
\begin{align}
    \mathcal{I}=\mathbf{Q}\mathbf{D}\mathbf{Q}^\top,\label{eq:eigendecomposition}
\end{align}
where the columns of $\mathbf{Q}$ constitute the orthonormal eigenvectors of $\mathcal{I}$ (our eigenmodes of the spectrum) and $\mathbf{D}$ is the diagonal matrix of eigenvalues, ordered by decreasing eigenvalues. The numerical construction is described in Sec.~\ref{sec:methods_eigenbasis}.

As the entries of $\mathbf{D}$ are eigenvalues of the information matrix $\mathcal{I}$, mode $i$ has inverse variance (in the Laplace approximation) given by the $i$-th diagonal element of $\mathbf{D}$. We thus construct a series of constrained spectral modes for eigenvalues above a certain threshold (or equivalently, variances below a particular threshold). To set a natural scale for comparing these eigenvalues, we define the mode amplitudes $a_i$ by
\begin{align}
    \mathbf{a} \equiv \mathbf{Q}^\top \boldsymbol{\theta}.
\end{align}

As shown in Fig.~\ref{fig:goodfig}, the most constrained (zeroth-order) mode acts as a bandpass-weighted flux normalisation term. The amplitude of this mode provides a scale against which the variances of the remaining modes may be compared. We define an unconstrained mode as any mode whose standard error (i.e. inverse square root of the information) exceeds a certain fraction of this zeroth-order mode.
\begin{align}
    \frac{1}{|a_0| \sqrt{d_i}} \geq \rho,\label{eq:constrained}
\end{align}
where $d_i$ is the information eigenvalue associated with mode $i$, $a_0$ is the zeroth element of the mode amplitude vector $\mathbf{a}$ (i.e. the zeroth mode coefficient), and $\rho$ is our choice of uncertainty tolerance. In Secs.~\ref{sec:fluxfov},~\ref{sec:read_sampling}, and~\ref{sec:jwst_hubble}, we adopt $\rho=0.1$. The choice of tolerance is discussed in Sec.~\ref{sec:methods_eigenbasis}.

After choosing an appropriate $\rho$, we recover $N$ constrained modes. We convert the number of constrained modes to an approximate spectral resolution $R$ as follows
\begin{align}
    R \approx \frac{\lambda_{\mathrm{cen}}}{(\lambda_{\max} - \lambda_{\min})/N},\label{eq:spectralRes}
\end{align}
where $\lambda_{\mathrm{cen}}$, $\lambda_{\max}$, and $\lambda_{\min}$ represent the central, maximum, and minimum wavelength transmitted by a particular bandpass, respectively (see Sec.~\ref{sec:methods_eigenbasis}). As this estimate is only approximate, the precise location of the reference wavelengths (e.g. arithmetic or geometric midpoint for $\lambda_{\mathrm{cen}}$) is not consequential; in the calculations below we take $\lambda_{\mathrm{cen}}$ as the arithmetic midpoint.

\subsubsection{Eigenmodes of the Polychromatic Airy Disk}\label{sec:airy_example}

We now examine a canonical example of a beam passing through a clear, circular aperture which spans an octave of bandwidth. We implement a Poissonian likelihood over a field of view of $10\times10$ $\lambda_{\min}/D$, with a total flux of $10^6$ photons and equal photon flux per sampled wavelength. The simulation is described in Sec.~\ref{sec:methods_airy}.

\begin{figure}
    \centering
    \makebox[\linewidth][c]{%
        \includegraphics[width=5.25in]{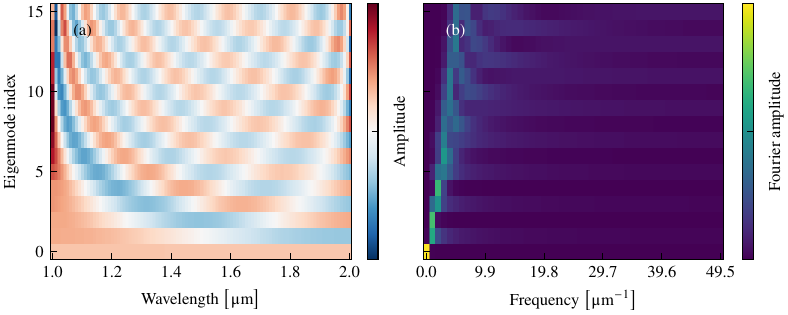}%
    }
    \caption[Spectral information eigenmodes]{Eigenmodes of the spectrum in wavelength space (a) and spectral-frequency space (b). Note that here `spectral frequency' means frequency of the spectral mode, not any individual temporal frequency or spatial wavenumber of a particular wavelength.}
    \label{fig:goodfig}
\end{figure}

The eigenmode with index zero (i.e., highest eigenvalue) is approximately flat across the relevant wavelength range, (see Fig.~\ref{fig:goodfig}(a)), with nearly all of its power concentrated in the zero-frequency band (see Fig.~\ref{fig:goodfig}(b)). The subsequent eigenmodes instead generally approximate bandpass-attenuated sinusoids, with each following eigenmode introducing an extra half-wave across the relevant bandpass. Figure~\ref{fig:goodfig}(b) demonstrates this marginal frequency increase per mode, and clarifies that these modes are not exact sinusoids, instead broadening in frequency with increasing order. This spectral broadening is expected, given that the spectral transfer function kernel found in Sec.~\ref{sec:spectralconv} operates in log-spectral frequency rather than (linear) spectral frequency.

\begin{figure}
    \centering
    \makebox[\linewidth][c]{%
        \includegraphics[width=5.25in]{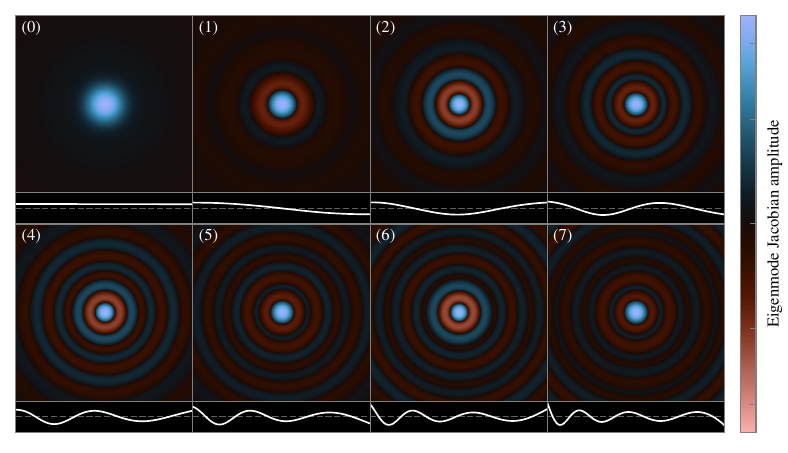}%
    }
    \caption[Image responses to spectral eigenmodes]{Jacobians of eigenmodes 0-7 of the Airy disk, with the corresponding eigenmode plotted below each image. Note the colour bar and fourth-root scaling.}
    \label{fig:eigmodejacobian}
\end{figure}

The eigenmode Jacobians, which describe the data response to small changes in the corresponding mode coefficients, are displayed in Fig.~\ref{fig:eigmodejacobian}. Here, it is more apparent that the mode $i$ has $i$ nodes (zero-crossings), and that the image-plane response to the odd and even eigenmodes differs substantially. In particular, it appears that Jacobian amplitudes alternate in sign radially, with odd and even eigenmodes exhibiting regions of broader (e.g. 2) and thinner (e.g. mode 3) periodic annuli. Note that the absolute sign of the Jacobian responses is arbitrary; here we choose it such that the response is positive in the centre. The first unconstrained mode, defined by Eq.~\eqref{eq:constrained}, occurs well before the eigenvalues approach numerical precision and the eigenmode structure deteriorates. This numerical behaviour is discussed in Sec.~\ref{sec:methods_eigenbasis}.

\subsubsection{Flux and FOV Scaling}\label{sec:fluxfov}

\begin{figure}
    \centering
    \makebox[\linewidth][c]{%
        \includegraphics[width=5.25in]{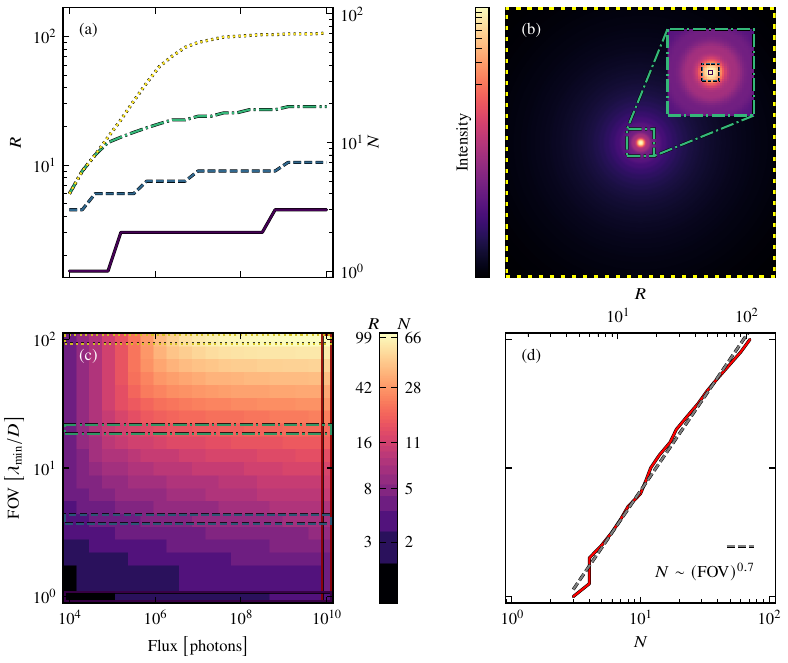}%
    }
    \caption[Flux and field-of-view dependence of spectral resolution]{Constrained modes $N$ and subsequent spectral resolution $R$ dependence on flux and field of view. (a) $N$ and $R$ dependence on flux for a selected sample of fields of view. (b) Selected fields of view (boxes delimited by 4 different linetypes) in the focal plane. Inset is an enlargement of the PSF core. (c) $N$ and $R$ values for a matrix of logarithmically spaced fluxes and fields of view. (d) $N$ and $R$ scaling and fitted power law with field of view for the maximum sampled flux value of $10^{10}$ photons. Note that the horizontal and vertical axes of (a) and (d) respectively share their values with (c).}
    \label{fig:4plots}
\end{figure}

Here we determine the joint flux-FOV scaling of the number of constrained modes $N$ and subsequently the spectral resolution $R$ in the canonical (flat octave bandpass, Airy disk) example. The spectral resolution map is shown in Fig.~\ref{fig:4plots}(c), with the calculation described in Sec.~\ref{sec:methods_fluxfov}. Note that the aperture and bandpass choice remain identical to Sec.~\ref{sec:airy_example}.

The number of constrained modes and thus spectral resolution increases with both incident flux and field of view; however, two additional trends are apparent. First, for a fixed field of view, the growth in the number of constrained modes diminishes with increasing flux, as shown in Fig.~\ref{fig:4plots}(a). That is, while the fractional uncertainty of each mode decreases as the inverse square root of flux (in the Poisson case), additional spectral degrees of freedom inside a particular FOV appear more difficult to reach. Second, Fig.~\ref{fig:4plots}(d) shows that in the high-flux limit the number of constrained modes increases approximately as a power law in field of view over the sampled range, giving
\begin{align}
    N \propto \left( \mathrm{FOV} \right)^{0.7}.
\end{align}

This sublinear empirical relation describes the effective rank of the spectral expected information matrix as dictated by our cutoff criterion over the sampled field-of-view range.

\subsubsection{Read Noise and Sampling Effects}\label{sec:read_sampling}

\begin{figure}
    \centering
    \makebox[\linewidth][c]{%
        \includegraphics[width=5.25in]{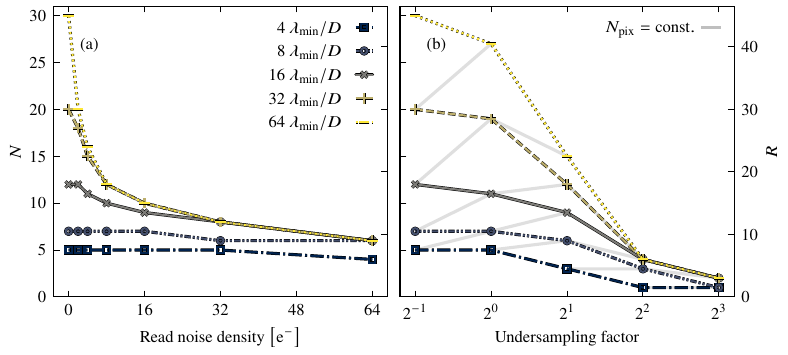}%
    }
    \caption[Read-noise and sampling effects on spectral resolution]{Number of constrainable modes $N$ and spectral resolution $R$ attainable at a range of fields of view, as functions of (a) read noise density, and (b) undersampling factor, for a circular aperture and flat octave bandpass, with a fixed flux of $10^6$ photons. Note that a Poissonian likelihood was used in (b), while (a) employed the Gaussian approximation with per-pixel variance equal to the sum of the Poisson shot-noise and detector read-noise variances.}
    \label{fig:read_nyquist}
\end{figure}

The recoverability of the spectral modes in any physical case is necessarily influenced by detector effects. Two notable considerations are detector readout noise and sampling. With the same circular aperture and octave bandpass configuration, we find that for a range of fields of view, increasing the readout noise density produces an effective FOV truncation, with low-intensity outer diffraction features dominated by this readout uncertainty as shown in Fig.~\ref{fig:read_nyquist}(a).

Conversely, the effects of undersampling are more nuanced. While for any fixed angular field of view (and here, in the absence of detector readout noise), sampling at a poorer rate reduces the recoverable modes and effective spectral resolution, when the number of detector pixels is held constant apparent gains can be made. That is, if the field of view is increased by increasing the angular pixel size (with the number of pixels remaining unchanged), the number of constrainable modes can increase, potentially up to a factor of $\approx 2$, shown in Fig.~\ref{fig:read_nyquist}(b). While the optimal undersampling factor for a fixed detector format will naturally depend on the signal-to-noise ratios of the outer diffraction features, the presence of this trade-off demonstrates that sampling and field of view should not be considered independently; under a fixed detector pixel count, better-than-Nyquist sampling is not necessarily ideal for recovery of spectral information. The detector-noise and sampling calculations are described in Sec.~\ref{sec:methods_read_sampling}.

\subsubsection{Fiducial Cases: \textsl{Webb}/NIRCam and \textsl{Hubble}/WFC3}\label{sec:jwst_hubble}

Here we investigate the spectral resolution scaling with the number of integrations for simulated imaging with real instruments and filters. We incorporate Poisson and detector-readout variance, with the source flux chosen such that the brightest pixel reaches the adopted below-saturation photoelectron count in each integration. The instrument configurations and likelihoods are described in Secs.~\ref{sec:methods_wfc3} and~\ref{sec:methods_nircam}.

\begin{figure}
    \centering
    \makebox[\linewidth][c]{%
        \includegraphics[width=5.25in]{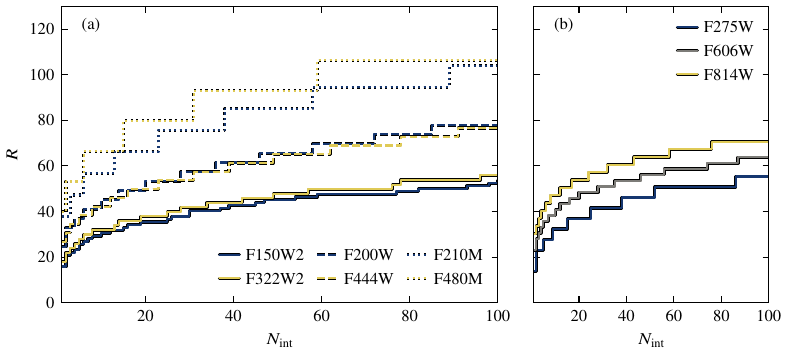}%
    }
    \caption[Spectral resolution forecasts for Webb and Hubble]{Spectral resolution $R$ scaling with number of integrations $N_{\mathrm{int}}$ for (a) representative \textsl{Webb}/NIRCam short- and long-wavelength channel filters and (b) three \textsl{Hubble}/WFC3 UVIS wide-band filters.}
    \label{fig:filter_Rs}
\end{figure}

In the \textsl{Hubble}/WFC3 case, we examine three representative wide-band filters using the $512\times512$ pixel subarray `UVIS2-M512C-SUB'. The results are displayed in Fig.~\ref{fig:filter_Rs}(b).

As expected, gains in spectral resolution $R$ diminish with increasing integrations, much like cases with fixed fields of view in Fig.~\ref{fig:4plots}(a). Notably, all three filters are significantly undersampled, with F275W particularly so. Although the attainable spectral resolution also depends strongly on the fractional bandwidth of the filter $\Delta \lambda/\lambda$ as shown in Fig.~\ref{fig:filter_Rs}(a), its severe undersampling appears to substantially limit the number of constrainable modes. The comparatively narrower fractional bandwidth partially compensates for this reduction when expressed as $R$, however it still remains below that of the longer-wavelength filters.

\begin{figure}
    \centering
    \makebox[\linewidth][c]{%
        \includegraphics[width=5.25in]{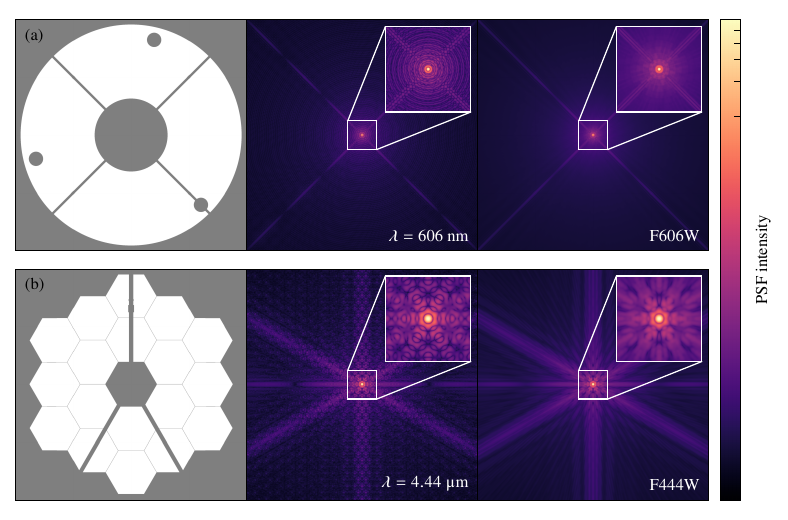}%
    }
    \caption[Hubble and Webb apertures and PSFs]{\textsl{Hubble} and \textsl{Webb} apertures and relevant PSFs. (a) The \textsl{Hubble}/WFC3 aperture (left), monochromatic PSF across the entire subarray field of view (middle), and the corresponding polychromatic PSF in F606W (right). (b) The \textsl{Webb}/NIRCam aperture (left), monochromatic PSF in the long-wavelength channel across the entire field of view, and the corresponding polychromatic PSF in F444W (right).}
    \label{fig:hubble_webb_psfs}
\end{figure}

In the \textsl{Webb}/NIRCam case, we examine representative short- and long-wavelength channel filters using the SUB400P $400\times400$ pixel point-source imaging subarrays. Note that the short- and long-wavelength channels have different plate scales, such that the longest-wavelength filters of each are partially Nyquist sampled. The spectral resolutions are displayed in Fig.~\ref{fig:filter_Rs}(a). The apertures and their corresponding monochromatic and polychromatic PSFs are displayed across the relevant fields of view in Fig.~\ref{fig:hubble_webb_psfs}.

The narrower medium-band filters in both channels significantly outperform the wider-band filters in terms of attainable spectral resolution. This is expected both from the direct dependence of $R$ on fractional bandwidth and from the characteristic bandwidth smearing induced on the PSF. Integrating over a broader bandpass reduces the contrast of its spatial features, thereby decreasing the sensitivity of the image to changes in the underlying spectrum and, consequently, the information available to constrain spectral modes.

\subsubsection{Spectral Injection and Recovery}\label{sec:injec_recov}

\begin{figure}
    \centering
    \makebox[\linewidth][c]{%
        \includegraphics[width=5.25in]{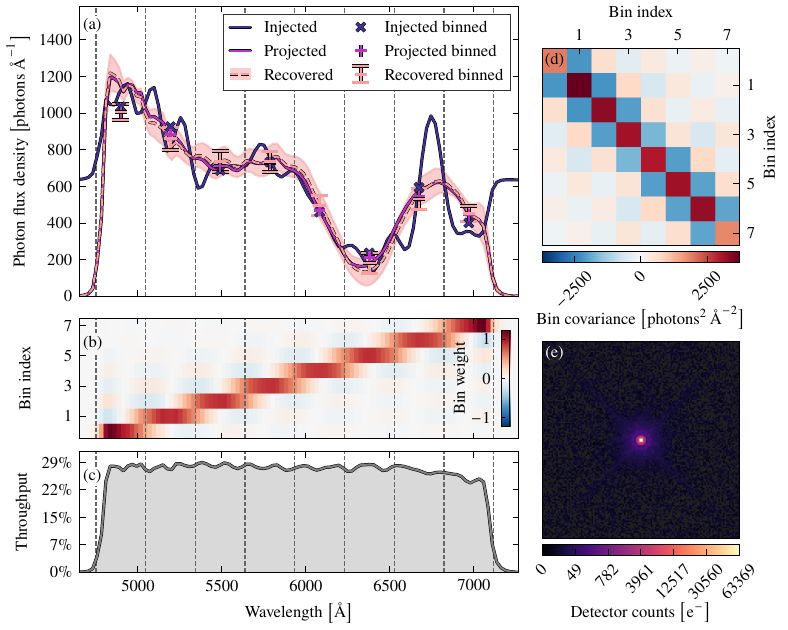}%
    }
    \caption[Spectral injection and recovery]{Spectral injection and recovery onto the retained modal basis. (a) Injected spectrum, its projection onto the recoverable spectral modes, and the recovered spectrum with its marginal $1 \, \sigma$ uncertainty. Also displayed are the corresponding quantities after integration into the equal-width bins, with boundaries indicated by the dashed vertical lines. (b) Effective bin weightings obtained by projection of the ideal top-hat bins onto the retained subspace, with oscillatory non-local components arising from the modal truncation. (c) Instrument throughput across the wavelength range. (d) Covariance matrix of the recovered bins. (e) Simulated polychromatic \textsl{Hubble}/WFC3 UVIS2 PSF from which the spectrum is recovered. Note the fourth-root scaling in displayed photoelectron count, with negative values displayed as grey.}
    \label{fig:inject_recov}
\end{figure}

We simulate a single integration of a \textsl{Hubble}/WFC3 UVIS2 observation, with an injected spectrum of 30 modes. We constrain 8 of these modes, with the procedure described in Sec.~\ref{sec:methods_recovery}. The results of this injection and recovery are displayed in Fig.~\ref{fig:inject_recov}.

The injected spectrum, its projection onto the recoverable spectral modes, and the recovered spectrum with its marginal $1\,\sigma$ uncertainty are compared in Fig.~\ref{fig:inject_recov}(a). Importantly, the propagated covariance represents uncertainties in the spectrum projected into the truncated modal basis, rather than uncertainties in the full untruncated spectrum. The covariance calculation is given in Sec.~\ref{sec:methods_covariance}. As also described in Sec.~\ref{sec:methods_covariance}, binning the projected spectrum is mathematically equivalent to applying effective bin weightings to the unprojected spectrum directly. The propagated covariance can therefore be interpreted as the uncertainty in these effective bins, which are approximate top-hats with minor oscillating non-local components, as illustrated in Fig.~\ref{fig:inject_recov}(b).

The bin intervals were optimised as described in Sec.~\ref{sec:methods_binning}. For the throughput curve shown in Fig.~\ref{fig:inject_recov}(c), this optimisation only produces a modest improvement over adopting the nominal bandpass limits directly. The bin covariance matrix and the simulated data are displayed in Fig.~\ref{fig:inject_recov}(d) and~(e), respectively. Importantly, the covariance matrix is mostly diagonal as the recovered bin amplitudes are only weakly correlated. With the locality of the weightings, the bins therefore provide a largely local representation of the recovered spectrum.

\section{Discussion and Conclusions}
\label{sec:discussion}
We first compare our method with previous approaches to spectral recovery from polychromatic PSFs. Taking the singular-value decomposition of the model matrix as described in Wang and Menon~\cite{wang2014b} is indeed similar to the procedure described above. In the limiting case of independent, uniform-variance Gaussian noise with a linear model, our information matrix decomposition would coincide with their SVD approach. As demonstrated in Nordebo \textsl{et al.}~\cite{nordebo2013regular}, singular-value decomposition of the forward operator is equivalent (up to scaling factors) to eigendecomposition of the relevant information matrix under the uniform-variance Gaussian likelihood. The major difference between our method and that of Wang and Menon's lies in the inversion process; Wang and Menon use Tikhonov regularisation rather than modal truncation, although truncating the singular values is discussed. Furthermore, while their method requires a specialised spectroscopic instrument, our results are generalisable to arbitrary diffractive PSFs and likelihoods with more complex optical and detector models.

The studies by Chen, Gu, and Liu~\cite{chen2024} and Cao \textsl{et al.}~\cite{cao2026} provide the closest comparison to our spectral information and recovery formalism. However, the number of spectral modes retained in these images is left unexplored; our eigendecomposition provides effective spectral resolution metrics while also permitting information-optimal recovery. We also demonstrate that our nonlocal modes can be projected to an approximately local top-hat-like basis, validating this measure of spectral resolution.

While these methods are applicable across almost any domain in optical imaging, they are perhaps most applicable in astronomy. The \textsl{Webb}/NIRCam and \textsl{Hubble}/WFC3 simulations demonstrate PSF spectroscopy under realistic noise considerations, implying that imaging data contains spectral structure beyond a single photometric observable. Considering the widespread availability of broadband PSFs in archival data (e.g., observations in Refs.~\citenum{pope2013, calissendorff2023}), our methods could enable recovery of low-resolution spectra across large numbers of targets without dedicated spectroscopic observation. This spectral recovery would involve marginalisation over more than just spectral parameters (for example source positions and the astrophysical scene, wavefront aberrations and other aperture-plane considerations). However, our information-theoretic formalism is well-suited to this task; one could evaluate the matrices across all model parameters to determine their covariance.

Correctly accounting for PSF chromaticity is necessary even when the spectrum is uninteresting to the observer. Astrometry, photometry, and high-contrast/coronagraphic imaging all necessarily require precise PSF calibration, with applicability simply contingent on the construction of a differentiable forward model for computing information matrices (e.g., Refs.~\citenum{feng2025, cao2026}). Furthermore, spectral coupling to the wavefront during broadband defocused wavefront sensing could present another worthwhile application (e.g. in \textsl{The Lazuli Space Observatory}~\cite{lazuli2026}). In these instances, fixing the spectrum to an assumed template rather than inferring the spectral content present in the observation may strongly bias the results, especially if the spectrum is significantly covariant with other parameters of interest. Constructing joint eigenbases of the spectrum and other parameters (for example, a joint spectrum-wavefront eigenbasis) and subsequent analysis of the covariance would be a natural extension of this work. With the ABCD matrix formalism for paraxial propagation and the differentiability of \jax and \dlux, the flow of spectral information as light propagates through arbitrary optical systems \cite{fisher_flow} may also be quantified: a task that we leave for future studies.

We again emphasise that the number of constrained modes $N$ and associated spectral resolution $R$ depend on the choice of the constrained mode cutoff $\rho$. While for our purposes, $\rho \approx 0.1$ presented a relatively small projected wavelength-level variance, introducing other covariant parameters may substantially change this outlook. In these cases, the cutoff may be flexibly adjusted to achieve any desired trade-off between spectral resolution and uncertainty. Furthermore, the information formalism for a Poissonian likelihood (unlike the fixed-covariance Gaussian likelihood with a linear model used in Sec.~\ref{sec:methods_recovery}) is inherently local; changes in the PSF will subtly affect the information matrix and subsequent eigenmodes. The extent of this effect is expected to increase with greater nonlinearities and distances from maximum likelihood. Precisely quantifying these eigenmode deviations was not attempted here and is left for future studies. Regardless, when inferring the spectrum in these regimes, one may iteratively reconstruct and truncate the eigenmodes as necessary while approaching maximum likelihood.

We have thus robustly quantified the spectral information content in diffraction-limited images and provided a method for its recovery. We have constructed the spectral transfer function, describing how spectral structure is transmitted, and have formalised the spectral information eigenbasis, quantifying what remains recoverable under observational constraints. We have further demonstrated that these modes and their uncertainties can be transformed into interpretable, localised spectral bins, with wide-ranging applications in both data processing from existing optical systems and experimental design in future instruments.

\section{Methods}\label{sec:methods}
Here we provide the relevant mathematical derivations, describe the spectral eigenmode construction, detail the parameters of the optical forward models, and present the spectral recovery procedure.

\subsection{Derivation of the Spectral Convolution}\label{sec:methods_spectralconv}

To prove Proposition~1, we start from Eq.~\eqref{eq:poly_g_int} and use the logarithmic coordinates $\eta=\log r$ and $\ell=\log\lambda$, with $d\lambda=e^\ell d\ell$. This transformation gives

\begin{align}
    p(e^\eta, \phi | S) = \int_{-\infty}^\infty S (e^\ell) e^{-\ell} g (e^{\eta - \ell}, \phi ) d\ell.
\end{align}

To ensure well-behaved Fourier kernels near $\eta=-\infty$ ($r=0$), we multiply both sides by $e^\eta$. Using the definitions in Eq.~\eqref{eq:log_functions}, we obtain
\begin{align}
    \Pi_\phi(\eta)=\int_{-\infty}^\infty\Sigma(\ell)K_\phi(\eta-\ell)d\ell,
\end{align}
which is the convolution in Eq.~\eqref{eq:spectral_conv}.

Under the assumptions of Proposition~1, $\Sigma$ and $K_\phi$ are integrable: in particular,
\begin{align}
    \int_{-\infty}^\infty|\Sigma(\ell)|d\ell
    &=\int_{\lambda_{\min}}^{\lambda_{\max}}\frac{|S(\lambda)|}{\lambda}d\lambda<\infty,\nonumber\\
    \int_{-\infty}^\infty|K_\phi(\eta)|d\eta
    &=\int_0^\infty|g(\xi,\phi)|d\xi<\infty.
\end{align}
The convolution theorem therefore gives Eq.~\eqref{eq:spectral_conv_fourier}, using the one-dimensional Fourier convention
\begin{align}
    \widehat{f}(\omega)=\mathcal{F}\{f\}(\omega)=\int_{-\infty}^\infty f(x)e^{-i\omega x}dx.\label{eq:fourier_convention}
\end{align}

We can further analyse $\widehat{\Sigma}(\omega)$ and $\widehat{K}_\phi$. By our definitions of $\ell$ and $\lambda$, we can write $\widehat{\Sigma}$ as
\begin{align}
    \widehat{\Sigma} (\omega) &= \int_{-\infty}^\infty S(e^\ell) e^{-i \omega \ell} d\ell, \nonumber \\ &= \int_{0}^\infty S (\lambda) \lambda^{-i \omega-1} d \lambda, \nonumber \\
    &= \mathcal{M} \{ S \} (-i \omega),
\end{align}
where $\mathcal{M}$ denotes the Mellin transform \cite{mellin_book}, defined by
\begin{align}
    \mathcal{M} \{ f\} (s) \equiv \int_0^\infty f(x) x^{s-1} dx.
\end{align}

With $\widehat{K}_\phi$, we find a similar Mellin integral
\begin{align}
    \widehat{K}_\phi(\omega)
    &=\int_{-\infty}^\infty e^\eta g(e^\eta,\phi)e^{-i\omega\eta}d\eta,\nonumber\\
    &=\int_0^\infty g(\xi,\phi)\xi^{-i\omega}d\xi
    =\mathcal{M}\{g_\phi\}(1-i\omega),
\end{align}
establishing Eq.~\eqref{eq:mellin_psf_kernel}. Here the reason for multiplication by $e^\eta$ becomes apparent; if omitted, the radial integral would contain an extra factor of $1/\xi$, introducing a singularity at $\xi=0$ for profiles with a nonzero central value.

We can express the original polychromatic PSF in its native coordinate vector $\mathbf{r}$ as
\begin{align}
    p(\mathbf{r}|S) &= \frac{1}{r} \mathcal{F}^{-1} \left\{ \widehat{\Sigma}(\omega) \cdot \widehat{K}_\phi (\omega) \right\} (\log r),
\end{align}
where $\mathcal{F}^{-1}$ denotes the inverse Fourier transform, defined by
\begin{align}
    f(x) = \mathcal{F}^{-1} \{ \widehat{f} \} (x) = \frac{1}{2\pi} \int_{-\infty}^\infty \widehat{f}(\omega) e ^{i\omega x} d\omega.
\end{align}

The apparent singularity at $r=0$ is an artefact of the log-radial representation, and the value at $r=0$ is understood by continuity.

To obtain the effective-resolution conversion in Eq.~\eqref{eq:log_resolution}, recall that adjacent extrema of a Fourier mode are separated by $\Delta\ell=\pi/|\omega|$. From $\ell=\log\lambda$, the corresponding wavelength interval centred geometrically on $\lambda$ is

\begin{align}
    \Delta \lambda &= \lambda e^{+\Delta \ell/2} - \lambda e^{-\Delta \ell /2}, \nonumber \\
    &= 2 \lambda \sinh (\Delta \ell /2).
\end{align}

Inverting and multiplying by $\lambda$ gives $R_{\mathrm{eff.}}=\lambda/\Delta\lambda$, as in Eq.~\eqref{eq:log_resolution}. The approximation follows from $\sinh x\sim x$ for small $x$.

\subsection{Derivation of the Airy Disk Spectral Transfer Function}\label{sec:methods_airykernel}

To prove Corollary~1, we use the Airy profile in Eq.~\eqref{eq:airy_g}. Azimuthal symmetry makes the spectral kernel independent of $\phi$, and its Fourier transform is

% Source: ms(5).tex L1019–1021
\begin{align}
    \widehat{K}(\omega) = \int_{-\infty}^\infty e^\eta \pi D^2 \jinc^2 (\pi D e^\eta) e^{-i\omega \eta} d\eta.
\end{align}

After significant algebraic manipulation, we find a Mellin transform of the square of the first-order Bessel function of the first kind
\begin{align}
    \widehat{K}(\omega) &= \pi^{i\omega} D^{1+i\omega} \int_0^\infty J_1^2(\beta)\beta^{-i\omega - 2} d\beta, \nonumber \\
    &= \pi^{i\omega} D^{1+i\omega} \mathcal{M} \{ J_1^2 \} (-i\omega - 1),
\end{align}
where we perform the change of variables $\beta \equiv \pi D e^\eta$.

The Mellin transform of $J_1^2$ is given by Eq.~\href{https://dlmf.nist.gov/10.22#E57}{(10.22.57)} of the DLMF \cite{dlmf}. Simplification with \textsl{Wolfram Mathematica}~\cite{Mathematica} gives Eq.~\eqref{eq:airy_spectral_kernel}.

For the asymptotic magnitude, the fixed-real-part Gamma-function asymptotic is given by Eq.~\href{https://dlmf.nist.gov/5.11#E9}{(5.11.9)} of the DLMF \cite{dlmf}:
\begin{align}
    |\Gamma(a+it)|\sim\sqrt{2\pi}|t|^{a-1/2}e^{-\pi|t|/2},\qquad |t|\rightarrow\infty.
\end{align}
Applying this to the Gamma factors in Eq.~\eqref{eq:airy_spectral_kernel}, with $t=\omega/2$, gives
\begin{align}
    |\widehat{K}(\omega)|\sim\frac{D}{2\sqrt{\pi}}\left|\frac{\omega}{2}\right|^{-5/2}
    =\frac{2\sqrt{2}D}{\sqrt{\pi}}|\omega|^{-5/2},
\end{align}
as stated in Eq.~\eqref{eq:airy_kernel_asymptotic}.

\subsection{Numerical Construction and Selection of Spectral Eigenmodes}\label{sec:methods_eigenbasis}

The method first involves approximating the polychromatic PSF given by Eq.~\eqref{eq:poly_g_int} by incoherently propagating a fixed number of wavelengths, weighted by an appropriate throughput function. After applying any relevant detector model (e.g. integrating over the pixel width in Secs.~\ref{sec:methods_read_sampling} and~\ref{sec:methods_wfc3}), we calculate the expected information matrix $\mathcal{I}$ with respect to the fluxes associated with each propagated wavelength (these correspond to our parameters $\boldsymbol{\theta}$).

The forecasts in Secs.~\ref{sec:airy_example}--\ref{sec:jwst_hubble} use expected information, which here corresponds to evaluating the Hessian at the expected image. The recovery procedure in Sec.~\ref{sec:injec_recov} instead evaluates observed information. We use \jax as applied through \dlux \cite{jax, dlux1} to differentiate the likelihood. 

The eigenmodes $\mathbf{q}_i$ are orthonormal under the Euclidean inner product. These eigenmodes are also by construction orthogonal under the information metric, with squared norm $\mathbf{D}_{ii}$, i.e., they satisfy
\begin{align}
    \mathbf{q}^\top_i \mathbf{q}_j &= \delta_{ij}, \nonumber \\
    \mathbf{q}^\top_i \mathcal{I} \mathbf{q}_j &= \mathbf{D}_{ij},
    \label{eq:orthogonality}
\end{align}
where $\delta_{ij}$ is the Kronecker delta function.

We adopt $\rho=0.1$ (Eq.~\eqref{eq:constrained}) for the forecasts, as above this value we found that the wavelength-space uncertainties associated with additional modes grew larger than the extra spectral structure retained. This choice is nonetheless flexible and may be tuned to meet a desired compromise between spectral resolution (from more modes) and reconstruction uncertainty (from greater modal uncertainty), especially if other (here unconsidered) parameters covaried with the spectrum. For examples and discussion of this information-eigenmode truncation, see Refs.~\citenum{tegmark1997, hunterer2003truncate, amigo2026, nordebo2013regular}.

The spectral resolution expression (Eq.~\eqref{eq:spectralRes}) follows from interpreting the number of recovered modes as the effective number of independently constrained spectral degrees of freedom across the bandpass. If the bandpass spans a width $\lambda_{\max}-\lambda_{\min}$, and is divided into $N$ spectral elements, the characteristic width of each element is $\Delta \lambda= (\lambda_{\max} - \lambda_{\min})/N$. Substituting this into the conventional definition of spectral resolution $R = \lambda/\Delta \lambda$ and evaluating at the central wavelength $\lambda_{\mathrm{cen}}$ yields the above expression.

While only the first 8 and 16 eigenmodes are displayed in Figs.~\ref{fig:eigmodejacobian} and~\ref{fig:goodfig} respectively, the total number of eigenmodes equals the number of wavelengths propagated (in our case, 101). However, for a given floating-point precision and signal-to-noise level (in the Airy example, a total flux of $10^6$ photons with Poissonian likelihood), later eigenmodes -- having eigenvalues within machine precision of zero -- instead appear as incoherent noise-like vectors with no discernible spectral structure. This breakdown in structure is an artefact of a particular choice of floating-point precision, as with greater precision the regular sinusoid-like structure of higher-order modes can be recovered. The near-zero eigenvalues imply that the modes are very poorly constrained, with the first unconstrained mode (defined by Eq.~\eqref{eq:constrained}) occurring well before their structure deteriorates.

\subsection{Simulated Apertures, Detectors, and Observations}\label{sec:methods_simulations}
Here we provide the parameters relevant to the inferred spectral resolutions of particular configurations and instruments.
\subsubsection{Canonical Airy Example}\label{sec:methods_airy}
We construct the relevant aperture and focal planes with the optical propagation and optimisation package \dlux, implementing a Poissonian likelihood over a field of view of $10 \times 10$ $\lambda_{\min}/D$. We set the total photon flux to $10^6$ photons, minimum wavelength $\lambda_{\min}$ to $1.0\times 10^{-6} \ \mathrm{m}$, and diameter $D$ to $1.0 \ \mathrm{m}$. We propagate 101 linearly spaced wavelengths, all with equal photon flux, and inspect the resulting eigenmodes.

Calculating and expressing the modes with uniform (rather than logarithmic) sampling in wavelength was chosen for easier physical interpretability of the resulting eigenmodes and consistency with the sampling conventions of instrumental throughputs and spectra in astronomy.

\subsubsection{Flux and Field-of-View Scaling}\label{sec:methods_fluxfov}
We calculate the expected information matrix and eigendecompose it for a range of logarithmically spaced fluxes and fields of view with a Poissonian likelihood, choosing $\rho=0.1$ as the constrained mode cutoff. The aperture and bandpass choice remain identical to Sec.~\ref{sec:methods_airy}. The resulting map is shown in Fig.~\ref{fig:4plots}(c), with the field-of-view power law fitted at the maximum sampled flux of $10^{10}$ photons, as shown in panel (d).
% add pupil sampling, or image oversampling?

\subsubsection{Read Noise and Sampling}\label{sec:methods_read_sampling}
We use a circular aperture and flat octave bandpass, with a fixed flux of $10^6$ photons. For the read-noise calculation in Fig.~\ref{fig:read_nyquist}(a), we employ the Gaussian approximation with per-pixel variance equal to the sum of the Poisson shot-noise and detector read-noise variances. For the sampling calculation in panel (b), we use a Poissonian likelihood without detector readout noise. We compare sampling changes at fixed angular field of view with changes at fixed detector pixel count; in the latter case, increasing the angular pixel size also increases the field of view. The detector model integrates over the pixel width. Read noise density is defined to be the RMS $e^{-}$ value per square Nyquist element (i.e. $(\lambda_{\min} / 2D)^2$ patch). Undersampling factor is defined to be the factor undersampling the Nyquist interval. A factor of 1 implies $\lambda_{\min}/2D$ (i.e. Nyquist) sampling, a factor of 2 implies $\lambda_{\min}/D$ sampling, a factor of 1/2 implies $\lambda_{\min}/4D$ sampling etc.

\subsubsection{\textsl{Hubble}/WFC3}\label{sec:methods_wfc3}
Noting that the expected information scales additively with the number of repeated experiments, we define an appropriately formulated Gaussian likelihood incorporating Poisson and detector-readout variance, evaluated at a source chosen such that the brightest pixel reaches the recommended maximum photoelectron count for the relevant detector formats.

In the \textsl{Hubble}/WFC3 case \cite{wfc3_paper}, we choose the field of view given by the $512 \times 512$ pixel subarray `UVIS2-M512C-SUB' for point source imaging, and compute the expected information matrix for a single exposure with the brightest pixel at $63 \ 000 \ \mathrm{e}^{-}$. We incorporate a constant readout noise of RMS $3 \ \mathrm{e}^{-}$ per pixel \cite{wfc3handbook}, and add the appropriate variances per pixel. For multiple integrations, this matrix is multiplied by the number of integrations $N_{\mathrm{int}}$. The number of constrained modes is computed by Eq.~\eqref{eq:constrained} and spectral resolution $R$ by Eq.~\eqref{eq:spectralRes}, for three representative wide-band filters. The results are displayed in Fig.~\ref{fig:filter_Rs}(b).

\subsubsection{\textsl{Webb}/NIRCam}\label{sec:methods_nircam}

The \textsl{Webb}/NIRCam case \cite{nircam_paper} incurs additional complications, due to the non-destructive up-the-ramp readout on HgCdTe infrared detectors. Here we adopt a single readout pattern of 10 groups, each consisting of a single frame with no skipped frames, corresponding to the `RAPID' readout pattern. We then propagate the per-frame readout noise through the ramp-fitting covariance as described in Ref.~\citenum{robberto2009}. Each integration is assumed to reach the recommended below-saturation flux in the brightest pixel ($100\ 000 \ \mathrm{e}^{-}$ and $78\ 000 \ \mathrm{e}^{-}$ in the short- and long-wavelength channels respectively) within the SUB400P $400 \times 400$ pixel point-source imaging subarrays \cite{nircam_docs}. Note that the short- and long-wavelength channels have different plate scales, such that the longest-wavelength filters of each are partially Nyquist sampled. Similarly to the WFC3 case, the expected information matrices were multiplied by the number of integrations $N_{\mathrm{int}}$ before the number of constrained modes were calculated and converted to spectral resolution $R$, displayed in Fig.~\ref{fig:filter_Rs}(a). The apertures and their corresponding monochromatic and polychromatic PSFs are displayed across the relevant fields of view in Fig.~\ref{fig:hubble_webb_psfs}.

\subsection{Spectral Recovery and Uncertainty Propagation}\label{sec:methods_inference}
Here we provide the methods used for the injection and recovery of a spectrum in a simulated broadband observation.

\subsubsection{Recovery Procedure}\label{sec:methods_recovery}

We perform spectral recovery by fitting the recoverable spectral modes. In this example, we initialise a flat spectrum with a total flux equal to the summed photoelectron count in the image, evaluate the observed information matrix at this spectrum, and diagonalise to construct orthogonal eigenmodes. The optimisation can then be restricted to the constrained modes as dictated by our selection criterion.

We use a $128 \times 128$ pixel cutout of \textsl{Hubble}/WFC3 UVIS2 detector, with the brightest pixel limit of $\approx 63,\!000 \ \mathrm{e}^{-}$, applying both Poisson shot noise and RMS $3 \ \mathrm{e}^{-}$ per pixel readout noise. We inject a spectrum of 30 modes, of which we constrain 8. As our model is linear in spectral eigenmodes with fixed Gaussian uncertainties (from summing read-noise and Poisson variances) the maximum-likelihood coefficients were obtained directly via a linear least-squares solve. We nonetheless tested use of first- and second-order optimisation with natural gradient descent and BFGS algorithms \cite{amari1998, broyden1970bfgs} to accommodate potential nonlinearities (for example, detector response functions) when applicable. The results of the injection and recovery simulation are displayed in Fig.~\ref{fig:inject_recov}.

In Sec.~\ref{sec:injec_recov} we choose a slightly stricter criterion, ignoring the last mode within the $\rho=0.1$ cutoff for improved visualisation of the recovered spectrum.

\subsubsection{Projected Covariance and Effective Bins}\label{sec:methods_covariance}

After truncating to the $N$ constrained modes, we propagate uncertainties to the projected spectrum in the original (wavelength-flux $\boldsymbol{\theta}$) basis via a linear transformation. Given the truncated $N\times N$ precision matrix $\mathbf{D}$ of eigenmode coefficients and the $N_\lambda \times N$ eigenmode matrix $\mathbf{Q}$ (where $N_\lambda$ is the original number of wavelength samples), the covariance matrix in the wavelength-flux basis $\mathbf{C}^{(\boldsymbol{\theta})}$ is given by
\begin{align}
    \mathbf{C}^{(\boldsymbol{\theta})} =  \mathbf{Q}_{(N)} \mathbf{D}_{(N)}^{-1} \mathbf{Q}_{(N)}^\top,
\end{align}
where the subscript $(N)$ indicates that we no longer consider the full $N_\lambda \times N_\lambda$ matrices and instead truncate to $N$ modes.

Importantly, this truncation only represents the covariance in the spectrum projected to those $N$ recovered eigenmodes \cite{hunterer2003truncate, nordebo2013regular}. It does not describe uncertainty in the full, unprojected spectrum.

Binning after modal truncation constitutes another linear transformation, and the covariance matrix in the binned fluxes $\mathbf{b}$ can be represented as
\begin{align}
    \mathbf{C}^{(\mathbf{b})} &= \mathbf{B}^\top \mathbf{C}^{(\boldsymbol{\theta})} \mathbf{B}, \nonumber \\
    &= \left(\mathbf{B}^\top \mathbf{Q}_{(N)} \right) \mathbf{D}^{-1}_{(N)} \left(\mathbf{B}^\top \mathbf{Q}_{(N)} \right)^\top, \label{eq:bin_covariance}
\end{align}
where $\mathbf{B}$ is the $N_\lambda \times N$ matrix whose columns represent the bin weightings (choosing the number of bins to be equivalent to the number of constrained modes $N$).

This covariance represents uncertainties in the bins of spectrum projected into the truncated modal basis, rather than the uncertainties in the bins of the true untruncated spectrum. However, the projection operator onto the retained modal subspace is
\begin{align}
    \mathbf{P} \equiv \mathbf{Q}_{(N)} \mathbf{Q}_{(N)}^\top, \label{eq:Pdef}
\end{align}
as the modes are Euclidean orthonormal (Eq.~\eqref{eq:orthogonality}). Since $\mathbf{P}$ is symmetric, we have
\begin{align}
    \mathbf{B}^\top \mathbf{P} \boldsymbol{\theta} = (\mathbf{P} \mathbf{B})^\top \boldsymbol{\theta}. \label{eq:effective_bin_projection}
\end{align}
Therefore, binning the projected spectrum is equivalent to applying the effective bin weightings $\mathbf{P} \mathbf{B}$ directly to the unprojected spectrum $\boldsymbol{\theta}$.

\subsubsection{Selection of the Binning Interval}\label{sec:methods_binning}

After recovering the spectrum and modal coefficients with the specified number of spectral modes, one can bin this spectrum into the same number of top-hat bins through left-multiplication by the transposed bin matrix $\mathbf{B}^\top$. The lower and upper limits of this bin matrix are jointly optimised to minimise the relative Frobenius-norm error $\epsilon_{\mathrm{bin}}$ between the ideal bin weightings $\mathbf{B}$ and their projections onto the retained modal subspace $\mathbf{P} \mathbf{B}$ (as defined in Eqs.~\eqref{eq:Pdef} and~\eqref{eq:effective_bin_projection})
\begin{align}
    \epsilon_{\mathrm{bin}} = \frac{\big\| \mathbf{P}\mathbf{B} - \mathbf{B}\big\|_\mathrm{F}}{\big\| \mathbf{B} \big\|_{\mathrm{F}}},
\end{align}
where $\|\cdot \|_{\mathrm{F}}$ denotes the Frobenius norm, defined as
\begin{align}
    \| \mathbf{A} \|_\mathrm{F} = \sqrt{\sum_i^m \sum_j^n |a_{ij}|^2},
\end{align}
where $a_{ij}$ is the $i,j$ element of the $m \times n$ matrix $\mathbf{A}$.

This criterion selects the wavelength interval where the effective bins most closely approximate the ideal bins after modal truncation, with these effective bins displayed in Fig.~\ref{fig:inject_recov}(b). Although for the throughput curve shown in Fig.~\ref{fig:inject_recov}(c) this optimisation only produces a modest improvement over adopting the nominal bandpass limits directly, it remains a principled way to generalise to throughput curves with broad or gradually decaying wings without well-defined cutoffs. For such bandpasses it may also be worthwhile to optimise the individual bin widths, however this would remove the simplicity and interpretability of uniform binning and is left for future studies.

\appendix    % this command starts appendixes

% \disclosures 
\subsection*{Disclosures}
The authors declare that there are no financial interests, commercial affiliations, or other potential conflicts of interest that could have influenced the objectivity of this research or the writing of this paper.

\subsection*{Acknowledgments}
% land acknowledgements
We acknowledge and pay respect to the traditional owners of the land on which the University of Sydney and Macquarie University are situated, upon whose unceded, sovereign, ancestral lands we work. We pay respects to their Ancestors and descendants, who continue cultural and spiritual connections to Country. 

% funding acknowledgements
BP and PT have been supported by the Australian Research Council grant DP230101439; ML and HG have been supported by the Australian Government Research Training Program (RTP) award. We are grateful to the Australian public for enabling this science. Development of \dlux has been supported by the Breakthrough Foundation through their Toliman project as a part of the Breakthrough Watch initiative. 

% software acknowledgements
This research made use of \textsc{NumPy} \cite{numpy}; Matplotlib \cite{matplotlib}; \textsc{Jax} \cite{jax}; \texttt{equinox} \cite{kidger2021equinox}; and \textsl{Mathematica} \cite{Mathematica}.

All code used to perform the analysis and produce the figures in this work is available at \url{https://github.com/milolangker/spectral-info}. 
% ai disclosure
Claude (Anthropic) and ChatGPT (OpenAI) were used for the development of some of the code. The authors take full responsibility for this work.

%%%%% References %%%%%

\bibliography{report}   % bibliography data in report.bib
\bibliographystyle{spiejour}   % makes bibtex use spiejour.bst

%%%%% Biographies of authors %%%%%

\vspace{2ex}\noindent\textbf{Milo Langker} is a PhD student in astronomy at the University of Sydney, where he previously studied for his undergraduate and Honours degrees. He has coauthored publications in optics and astronomy, with his current research interests including the information-theoretic limits of optical systems and their applications toward novel inference techniques.

\vspace{1ex}
\noindent Biographies of the other authors are not available.

\listoffigures
\listoftables

% for submission
% \end{spacing}
\end{document}